\documentclass[11pt, twoside]{article}
\usepackage{a4wide,amssymb,cite}
\usepackage{epsfig}
\usepackage[usenames,dvipsnames]{color}
\usepackage{slashed}

\usepackage{epsf,epsfig,subfigure,graphicx,amsmath,amssymb}
\usepackage{color}

\newcommand{\dis}[1]{\begin{equation}\begin{split}#1\end{split}\end{equation}}

\def\gsim{\mathrel{\rlap{\lower4pt\hbox{\hskip1pt$\sim$}}
    \raise1pt\hbox{$>$}}}                
\def\lsim{\mathrel{\rlap{\lower4pt\hbox{\hskip1pt$\sim$}}
    \raise1pt\hbox{$<$}}}                

\newcommand{\be}{\begin{equation}}
\newcommand{\ee}{\end{equation}}
\newcommand{\bea}{\begin{eqnarray}}
\newcommand{\eea}{\end{eqnarray}}

\begin{document}

\color{black}

\begin{flushright}
KIAS-P13009
\end{flushright}

\vspace{1cm}
\begin{center}
{\huge\bf\color{black}  General self-tuning solutions \\[3mm] and no-go theorem}\\
\bigskip\color{black}\vspace{1.5cm}{
{\large\bf Stefan F\"orste$^{a}$, Jihn E. Kim$^{b}$ and Hyun Min Lee$^{c}$}
\vspace{0.5cm}
} \\[7mm]

{\em $(a)$ {Bethe Center for Theoretical Physics and
Physikalisches Institut der Universit\"at Bonn, \\
Nussallee 12, 53115 Bonn, Germany.
}}\\
{\it $(b)$ Department of Physics and Center for Theoretical Physics,
Seoul National University, Seoul 151-747, Korea. } \\
{\it $(c)$ School of Physics, KIAS, Seoul 130-722, Korea.  }\\
\end{center}
\bigskip
\centerline{\large\bf Abstract}
\begin{quote}\large
We consider brane world models with one extra dimension. In the bulk
there is in addition to gravity a three form gauge potential or
equivalently a scalar (by generalisation of electric magnetic
duality). We find classical solutions for which the 4d effective
cosmological constant is adjusted by choice of integration
constants. No go theorems for such self-tuning mechanism are
circumvented by unorthodox Lagrangians for the three form respectively
the scalar. It is argued that the corresponding effective 4d theory
always includes tachyonic Kaluza-Klein excitations or ghosts. Known
no go theorems are extended to a general class of models with
unorthodox Lagrangians.

\end{quote}

\thispagestyle{empty}

\normalsize

\newpage

\setcounter{page}{1}

\section{Introduction}

Cosmological evolution of the universe is well described by the
Standard Big Bang Cosmology augmented with cold dark matter and
cosmological constant as input parameters, the so called $\Lambda {\rm
  CDM}$ model. However, the observed cosmological constant is
extremely small compared to the theoretical expectation with
Planck-scale cutoff (by a factor of about $10^{-120}$). Since there is
no symmetry prediciting vanishing cosmological constant without conflicting
observations, there must be a huge fine-tuning to cancel
contributions of different origin and size to the cosmological
constant at the observed level. This is the notorious cosmological
constant problem.

In this paper, we revisit the brane model with a three form field in
5d to obtain the flat space solution without a fine-tuning, namely,
the self-tuning solution.
The idea is that even if the cosmological constant on the brane, in
other words, the brane tension, is arbitrary, its effective 4d value
can be adjusted by choice of integration constants, i.e.\
without a fine-tuning \cite{Kachru:2000hf,ArkaniHamed:2000eg}.

However, there is a no-go theorem for such self-tuning solutions with
a canonical scalar field in 5d. In  detail, it has been shown that
either a naked singularity in the bulk or an infinite 4d Planck mass are
unavoidable \cite{Forste:2000ps,Forste:2000ft,Csaki:2000wz}. The first
two references consider particular models whereas the third provides
arguments valid for general scalar potentials. There were attempts of shielding the naked singularity of the self-tuning solutions with asymmetric warp factors  by the blackhole-like horizon \cite{asymwarp} but there is a no-go theorem associated with those too \cite{cline}. 
Instead of a canonical scalar field, an unorthodox action
with a three form field or a dual scalar has been considered and shown
to give rise to self-tuning solutions without naked singularity and
with a finite 4d Planck mass \cite{kkl,generalST,dSsolution,dualaction,pospelov}. (Later, unorthodox bulk
matter has been considered in a different approach \cite{antoniadis}.)

More recently, it has been pointed out \cite{stefan} that the original self-tuning
solutions with Lagrangian containing $1/H^2$, where $H^2$ is
constructed from the field
strength of the three form field, has tachyonic Kaluza-Klein (KK)
masses under the perturbation of the dual scalar through the bulk
space\footnote{See also Ref.~\cite{medved} for the earlier discussion on the stability of the self-tuning solution.}. Since there is no gap in the KK masses for
the infinite extra
dimension, the continuum of the tachyonic KK masses would pose a
serious problem for the stability of the self-tuning solution. The
motivation of this paper is to show whether the discovered instability
is generic for the self-tuning solutions.

For solving Einstein equations it proves useful to work with the bulk
theory given in terms of a three form gauge potential, whereas
the stability analysis is conveniently performed in its dual
formulation with a bulk scalar.
We first take an exponential type of the bulk Lagrangian, $e^{H^2}$, as an
alternative to the original self-tuning action with $1/H^2$.  In this
case, it is plausible to think of the bulk action as being derived
from the effective action of string compactifications. The vacuum
value of the field strength does not have to be nonzero in order to
satisfy the field equation,  so there might be a variety of the
cosmological solutions, for instance, cosmic inflation and graceful
exit might be described in the same setup.
 Furthermore, we can solve for a dual Lagrangian where a canonical
 kinetic term for the scalar is replaced by its
 Lambert W-function.
From the action quadratic in scalar perturbations, it is seen that the
self-tuning solution with $e^{H^2}$ Lagrangian has
 no 4d ghost but allows for tachyonic KK masses in some region of the
 bulk space. Then, we discuss the implication of the tachyonic KK
 masses for the cosmological constant problem in disguise.
Furthermore, considering a general form of the bulk action with a
three form field, we generalize the stability conditions for the
self-tuning solution and find that perturbations of the
self-tuning solution give rise to a ghost or tachyonic KK instability
for any continuous form of the bulk action.

The paper is organized as follows.
We begin with the general discussion on the self-tuning solution and
nearby curved solutions for a general form of the bulk action with a
three form field.
Then, choosing an example with Lagrangian $e^{H^2}$, we analyze the 
action quadratic in perturbations in the dualization. We also derive the
stability conditions of the quadratic action for a general form of the
bulk action and compare them to the self-tuning condition. Finally,
conclusions are drawn.

\section{Self-tuning model with a three form field}

Self-tuning solutions with a three form field \cite{kkl,generalST,dSsolution} and with a dual scalar \cite{dualaction,pospelov,stefan} have been considered
in the past. In this class of models the general action is \cite{generalST}
\begin{equation}\label{eq:action}
S = \int d^4x dr \sqrt{-g} \left\{ \frac{R}{2} - \Lambda_b + K\left(
    H^2\right)\right\} + \int_{r=r_0} d^4 x \sqrt{-g_4} \left(
    -\Lambda_1\right).
\end{equation}
We will parameterise the extra dimension by a `radial' coordinate
running from zero to infinity. The brane is positioned at $r =
r_0$. There are two cosmological constants in the 5d
theory: $\Lambda_b$ is the bulk constant whereas $\Lambda_1$ is the
brane constant. The constant $\Lambda_1$ contains effects of quantum
fluctuations on the brane which are considered as integrated out. So,
naturally, $\Lambda_1$ is given by the cut-off scale of the QFT living
on the brane, the 4d Planck mass for instance. Gravity is five
dimensional and in addition there is a three form potential $A_{NPQ}$
propagating in the bulk. Its $U(1)$ invariant field strength is $H$ and
in $H^2$ all tangent space indices are contracted, i.e.\
\begin{equation}
H_{MNPQ}= \partial_{[ M} A_{NPQ]} \,\,\, ,\,\,\, H^2 = H_{MNPQ}
H^{MNPQ} .
\end{equation}
In addition, a surface term is needed to cancel otherwise disturbing
terms in the variation of (\ref{eq:action}),
\begin{equation}\label{eq:surface}
S_{surf} = -2 \int d^4 x dr \partial_M \left( \sqrt{-g}
  \frac{ \partial K\left( H^2\right)}{\partial H^2}
  H^{MNPQ}A_{NPQ}\right) .
\end{equation}
Note, that this term vanishes if the expression under the partial
derivative is continuous across the brane and vanishes for $r=0$ and
$r=\infty$.

\subsection{Self-tuning solutions}

For the 5d metric we choose the ansatz
\begin{equation}\label{eq:5dmetric}
ds ^2 = a^2\left( r\right) \eta_{\mu\nu} dx^\mu dx^\nu +
\frac{dr^2}{f^2\left( r\right)} ,
\end{equation}
where $\eta_{\mu\nu} = \mbox{diag}\left( -1,1,1,1\right)$ denotes the
4d Minkowski metric. The Ricci tensor has the following non vanishing
components
\dis{
R_{rr} & =  -4 \frac{a^{\prime\prime}}{a} - 4 \frac{ a^\prime
  f^\prime}{af} , \\
R_{\mu\nu} & = -\left[ f^2 a a^{\prime\prime} + f f^\prime a a^\prime
  + 3 f^2 \left( a^\prime\right)^2\right] \eta_{\mu\nu} .
}
The equation of motion(eom) for the three form
potential reads
\begin{equation} \label{eq:H-eq}
\partial_M\left( \sqrt{-g} \frac{\partial K\left( H^2\right)}{\partial
    H^2}
  H^{MNPQ}\right) = 0.
\end{equation}
To respect 4d Lorentz invariance we impose that nothing depends on the
$x^\mu$, and $H^{r NPQ} =
0$. That is, the only non vanishing components are
\begin{equation}\label{eq:Hansatz}
H_{\mu\nu\rho\lambda} = \sqrt{ - \frac{a^8 H^2}{4!}}
\epsilon_{\mu\nu\rho\lambda} ,
\end{equation}
where $\epsilon_{\mu\nu\rho\lambda}$ is completely antisymmetric and
$\epsilon_{0123} =1$. The expression under the squareroot is not
negative due to the Minkowski signature.
Then equation (\ref{eq:H-eq}) is automatically satisfied.

The energy momentum tensor of the three form is
\begin{equation}
T^H_{MN} = K\left( H^2\right) - 8 \frac{\partial K\left(
    H^2\right)}{\partial H^2} H_{MPQR} {H_{N}}^{PQR} .
\end{equation}
With (\ref{eq:Hansatz}) we find for the non-vanishing components
\begin{equation}
T^H_{rr} = \frac{K\left( H^2\right)}{f^2} \,\,\, ,\,\,\,
T^H_{\mu\nu} = \left[  K\left( H^2\right) - 2 \frac{\partial K\left(
    H^2\right)}{\partial H^2} H^2\right] a^2 \eta_{\mu\nu} .
\end{equation}
Taking into account also contributions from bulk and brane
cosmological constants, the Einstein equations become
\begin{align}
R_{rr} - \frac{1}{2} g_{rr} R & \equiv 6 \frac{\left(a^\prime\right)^2}{a^2}
 = \frac{1}{f^2}\left( -\Lambda_b + K\left( H^2\right)\right)
 ,\label{eq:eirr}\\
R_{\mu\nu} - \frac{1}{2} g_{\mu\nu} R & \equiv\left[ 3 f^2 a
  a^{\prime\prime}
  + 3 f f^\prime a a^\prime +3f^2 \left(a^\prime\right)^2\right]
\eta_{\mu\nu} \nonumber\\ & = \left[ -\Lambda_b - \Lambda_1 \delta\left( r -
    r_0\right) \left| f\right| + K\left( H^2\right) - 2 \frac{\partial
    K\left( H^2\right)}{\partial H^2} H^2\right] a^2 \eta_{\mu\nu}
. \label{eq:eimunu}
\end{align}
The $rr$ component of Einstein's equation (\ref{eq:eirr}) can be used
to solve for the $rr$ metric component
\begin{equation}\label{eq:rrsol}
f^2 = \frac{a^2}{6\left( a^\prime\right)^2}\left( -\Lambda_b + K\left(
    H^2\right)\right) .
\end{equation}
This implies that the right hand side should be positive.
Now, we notice that our metric ansatz (\ref{eq:5dmetric}) allows for a
choice of $a\left( r\right)$ via redefinitions of the coordinate
$r$. Restrictions on $a$ arise when considering junction
conditions at the brane. For the induced metric on the brane to be
well defined, metric components $a^2$ and $f^2$ should be continuous at
$r=r_0$. That implies that the only candidate to create the
delta function singularity in (\ref{eq:eimunu}) is the term containing
$a^{\prime\prime}$. Hence, $a^\prime$ should jump when crossing the
brane. A choice for $a$ meeting that condition is
\dis{
&a\left( r\right) = \Theta\left( r_0 -r\right) r + \Theta\left( r-
  r_0\right) \frac{r_0 ^2}{r}\\
&a'(r)=\Theta(r_0-r)-\Theta(r-r_0)\frac{r_0^2}{r^2}. \label{4dwarp}
}
Indeed, we have a delta function singularity in $a^{\prime\prime}$,
\begin{equation}\label{eq:e}
a^{\prime\prime}= -2 \delta\left( r - r_0\right) + 2 \Theta\left( r -
  r_0\right) \frac{r_0^2}{r^3} .
\end{equation}
Matching the coefficients at the delta functions on left and right
hand sides of (\ref{eq:eimunu}) leads to the junction condition
\begin{equation}\label{eq:match}
\sqrt{ - \Lambda_b + K\left( H^2\right)}_{\left| r =r_0\right.} =
\frac{\Lambda_1}{\sqrt{6}}.
\end{equation}
For constant $K\left( H^2\right)$, which is actually the
Randall--Sundrum model, this is a fine tuning
condition. However, for non
trivial $K\left( H^2\right)$ the left hand side (lhs) can depend on integration
constants. Hence,
if equation (\ref{eq:match}) can be solved for integration constants
as a function of
the other parameters ($\Lambda_b, \Lambda_1$ and parameters appearing
in $K\left( H^2\right)$) no fine tuning is imposed.

Now, we solve the remaining non-singular part of (\ref{eq:eimunu}). It
is useful to rewrite (\ref{eq:rrsol}) as
\begin{equation}
f^2 = \frac{r^2}{6}\left( - \Lambda_b + K\left( H^2\right)\right) ,
\end{equation}
where we used that
$\Theta\left( x\right) + \Theta\left( -x\right)\equiv 1. $
Then we get
$$ 3 f f^\prime =
\frac{r}{2}\left( - \Lambda_b + K\left( H^2\right)\right) +\frac{r^2}{4}
  \frac{\partial K\left( H^2\right)}{\partial H^2} \frac{\partial
    H^2}{\partial r} . $$
With that it is not difficult to see that (\ref{eq:eimunu}) boils down
to (for $\partial K\left( H^2\right)/\partial H^2 \not= 0$)
\begin{equation}\label{eq:Hdif}
\left(  \Theta\left( r - r_0\right) -\Theta\left( r_0 -r\right)\right) r
  \frac{\partial H^2}{\partial r} = 8 H^2 .
\end{equation}
A continuous solution is
\begin{equation}
H^2 = -Q\left( \Theta\left(r_0 - r\right) r^{-8} + \Theta\left( r -
    r_0\right) r_0^{-16} r^{-8}\right) ,
\end{equation}
where $Q$ is a positive integration constant.
As a cross check we compute the effective 4d cosmological constant
which is given by
\begin{equation}
\Lambda_{4d} = \int_0^\infty dr \sqrt{-g}\left\{ \frac{R}{2}
  -\Lambda_b + K\left( H^2\right) - 2 \frac{\partial K\left(
      H^2\right)}{\partial H^2} H^2 \right\} - e^4\left( r_0\right)
\Lambda_1 ,
\end{equation}
where the penultimate term is the surface term (\ref{eq:surface}) with
equation of motion (\ref{eq:H-eq}) imposed. Plugging this in the metric anstatz
(\ref{eq:5dmetric}) leads to
\begin{align}
\Lambda_{4d} = & \int_0^\infty \frac{a^4}{\left| f\right|}\left\{ -4
  \frac{f^2 a^{\prime\prime}}{a} - 4 \frac{ f f^\prime a^\prime}{a} -
  6 \frac{f^2 \left( a^\prime\right)^2 }{a^2} -\Lambda_b + K\left(
    H^2\right) - 2 \frac{\partial K\left(
      H^2\right)}{\partial H^2} H^2 \right\}\nonumber \\
& - a^4\left( r_0\right)
\Lambda_1 .
\end{align}
Insertion of our solution (\ref{eq:rrsol}), (\ref{eq:e}) and
(\ref{eq:Hdif}) yields
\begin{align}
\Lambda_{4d} = & 8\int_0^\infty \delta\left( r- r_0\right)
r_0^4\sqrt{\frac{-\Lambda_b + K\left( H^2\right)}{6}} \nonumber \\
&+2 \int_0 ^{r_0}\left\{ -4 r^3 \sqrt{\frac{-\Lambda_b + K\left(
        H^2\right)}{6}}
  -\frac{r^4}{2} \frac{1}{\sqrt{6\left( -\Lambda_b + K\left(
          H^2\right)\right)}} \frac{\partial K\left(
      H^2\right)}{\partial H^2} \frac{\partial H^2}{\partial
    r}\right\}\nonumber \\ &
 - r_0 ^4 \Lambda_1 ,
\label{eq:Lcal}
\end{align}
where we substituted $r \to r_0^2/r$ in the region between $r_0$ and
infinity resulting in the factor of two in the second line of
(\ref{eq:Lcal}). Finally, we get
\begin{align}
\Lambda_{4d} & = 8 r_0^4 \sqrt{\frac{ -\Lambda_b + K\left(
      H^2\right)}{6}}_{\left| r = r_0\right.} -2 \int_0^{r_0}
  dr \partial_r\left( r^4 \sqrt{\frac{ -\Lambda_b + K\left(
      H^2\right)}{6}}\right) - r_0^4 \Lambda_1 \nonumber\\
& = 6  r_0^4 \sqrt{\frac{ -\Lambda_b + K\left(
      H^2\right)}{6}}_{\left| r = r_0\right.} -r_0^4 \Lambda_1
\nonumber \\
& = 0,
\end{align}
where we assumed that $K\left( H^2\right)$ is such that there is no
contribution at $r=0$ (or $r=\infty$) and used the junction condition
(\ref{eq:match}).

For the selftuning solution to be a non trivial candidate as a
solution to the fine tuning problem of the cosmological constant we
have to ensure that the 4d Planck mass is finite. Otherwise gravity
decouples and there is no backreaction of vacuum energy on
spacetime geometry. The 4d Planck mass is finite if the integral
\begin{equation}
I = \int_0^\infty dr \sqrt{-g} \,\frac{a^{2}}{f},
\end{equation}
relating the 5d Planck scale to the effective 4d Planck mass is finite.
With our solution this integral takes the form
\begin{equation}
I= 2 \int_0 ^{r_0}dr r \sqrt{\frac{6}{-\Lambda_b + K\left(
      H^2\right)}} = \int_0^{r_0^2} dx \sqrt{\frac{6}{-\Lambda_b + K\left(
      -Q/x^4\right)}}.
\end{equation}
Requiring finite Planck mass poses conditions on the function $K\left(
  H^2\right)$. An asymptotic AdS space at $a=0$ or $r=0$ is a sufficient condition for the self-tuning solution to give a finite 4d Planck mass, following the argument in Ref.~ \cite{kkl}:
\be
  a^2 K(H^2)\rightarrow 0\quad \quad{\rm for}\,\,\, a\rightarrow 0. \label{stcond}
\ee
Note, that
there are also other conditions coming from the requirement that
there should be no curvature singularities away from the brane's
position. For instance, the canonical choice $K\left( H^2\right) \sim
H^2$ is excluded.

\subsection{Nearby curved solutions}

Even if there is a self-tuning flat solution for a given form of the action, nearby curved solutions are generically allowed too.
In this section, we consider nearby curved solutions with nonzero 4D effective cosmological constant and the junction condition to be satisfied on the brane.

For 4D maximally symmetric solutions with warp factor, the Einstein equations, (\ref{eq:eirr}) and (\ref{eq:eimunu}), are modified to
\begin{align}
&  6 \frac{\left(a^\prime\right)^2}{a^2}-\frac{6\lambda}{a^2f^2}
 = \frac{1}{f^2}\left( -\Lambda_b + K\left( H^2\right)\right)
 ,\label{eq:eirr1}\\
& 3 f^2 a
  a^{\prime\prime}
  + 3 f f^\prime a a^\prime +3f^2 \left(a^\prime\right)^2-\frac{3\lambda}{a^2 f^2}
 \nonumber\\ & = \left[ -\Lambda_b - \Lambda_1 \delta\left( r -
    r_0\right) \left| f\right| + K\left( H^2\right) - 2 \frac{\partial
    K\left( H^2\right)}{\partial H^2} H^2\right] a^2
 \label{eq:eimunu1}
\end{align}
where $\lambda$ is defined from the 4D Ricci tensor, $R_{\mu\nu}=3\lambda g_{\mu\nu}$.
Then, the $rr$ component of Einstein equation (\ref{eq:eirr1}) can be solved for
\be
f^2=\frac{a^2}{6\left( a^\prime\right)^2}\left(\frac{6\lambda}{a^2} -\Lambda_b + K\left(
    H^2\right)\right) . \label{fsol}
\ee
Choosing the warp factor to be as in eq.~(\ref{4dwarp}), the similar junction condition as for the flat solution leads to
\be
\sqrt{\frac{6\lambda}{r^2}-\Lambda_b+K(H^2)}_{\left| r =r_0\right.}=\frac{\Lambda_1}{\sqrt{6}}.
\ee
Therefore, we get the 4D effective cosmological constant in terms of bulk and brane cosmological constants and the parameter in the solution of the anti-symmetric tensor field,
\be
\lambda=\frac{r^2_0}{6}\left[\frac{\Lambda^2_1}{6}-|\Lambda_b|-K(H^2)_{\left| r =r_0\right.}\right].  \label{effcc}
\ee

\section{Self-tuning with Lambert dynamics}

In this section, we take an example for the self-tuning solution and study the stability against fluctuations of the three form potential, before going into the general discussion in next section. In order to circumvent gauge fixing issues we will perform the calculation in a dual picture where the three form is replaced by a scalar.  Then, from the nearby curved solutions, we will also discuss the fine-tuning problem of obtaining the observed cosmological constant when an additional brane is introduced to cure the instability problem of tachyonic KK masses.

\subsection{The flat solutions}

The example takes the following form of the action \cite{generalST},
\begin{equation}
K\left( H^2\right) =-V \mbox{e}^{p H^2} , \,\,\, \mbox{with} \,\,\,
V>0,\,\,\, p>0.
\end{equation}
Then, from eq.~(\ref{eq:match}) with $H^2=-Q/r^4$, we obtain the junction condition for the brane with tension $\Lambda_1$ located at $r=r_0$ as
\be
V\, e^{-pQ/r^8_0} = |\Lambda_b| -\frac{\Lambda^2_1}{6M^3_5}.
\ee
Thus, there is a solution to the above junction condition for $|\Lambda_1|< \sqrt{6M^3_5 |\Lambda_b|}$. Otherwise, there is no flat solution satisfying the junction condition \footnote{There are similar results in the case with $K(H^2)=1/H^2$ too \cite{kkl}.}.
Next, we get
\begin{equation}
I = \int_0^{r_0^2}dx \sqrt{\frac{6}{-\Lambda_b - V\mbox{e}^{-pQ
      x^{-4}}}} \,\,\, \
\Rightarrow
\sqrt{\frac{6}{-\Lambda_b-V \mbox{e}^{-pQr_0^{-8}}}} < \frac{I}{r_0^2} <
  \sqrt{\frac{6}{-\Lambda_b}} , \label{eq:effplanck}
\end{equation}
and hence a finite effective Planck mass (note that $\Lambda_b$ has to
be negative for getting a non negative $f^2$ everywhere).

To construct the action for a dual scalar we start with \cite{dualaction,stefan}
\begin{equation}\label{eq:Hac}
S_H = -\int d^5 x \left\{ \sqrt{-g}\,V \mbox{e}^{p H^2} -\partial_M \phi
\epsilon^{MNPQR} H_{NPQR}\right\} ,
\end{equation}
where we view $\phi$ and $H_{NPQR}$ as independent fields. Varying
(\ref{eq:Hac}) w.r.t.\ $H_{NPQR}$
yields
\begin{equation}\label{eq:Hvar}
\partial_R \phi \epsilon^{RMNPQ} = 2\sqrt{-g} Vp H^{MNPQ}
\mbox{e}^{pH^2} .
\end{equation}
Computing the variational derivative w.r.t. $\phi$ we impose
boundary conditions on the scalar variation such that
$\delta\phi\epsilon^{r \mu\nu\rho \sigma}H_{\mu\nu\rho\sigma}$
vanishes at $r=0$ and $\infty$ and is continuous across the
brane. Then the $\phi$ equations of motion are
\begin{equation}
\partial_{\left[ M\right. } H_{\left. NPQR\right]} = 0.
\end{equation}
That is, $H_{MNPQ}$ can be written as a field strength of a three form
potential. The equation of motion (\ref{eq:H-eq}) follows from
(\ref{eq:Hvar}) and $\partial_M\partial_N \phi =\partial_N\partial_M
\phi$. So, (\ref{eq:Hac}) together with the discussed boundary
conditions on the scalar variation reproduces the action used in the
previous section. The dual formulation is obtained by eliminating $H$
via solving its algebraic equations of motion, i.e.\
\begin{align}
H^{MNPQ}& = \frac{1}{2\sqrt{-g}Vp}\,\mbox{e}^{-pH^2}\epsilon^{RMNPQ}\,\partial_R
\phi ,\\
H_{MNPQ} & = \frac{\sqrt{-g}}{2Vp}\,
\mbox{e}^{-pH^2}{\epsilon^{R}}_{MNPQ}\,\partial_R\phi ,\\
H^2 & = - \frac{4!}{\left(2Vp\right)^2}\mbox{e}^{-2pH^2}\left( \partial
  \phi\right)^2. \label{eq:Hasphi}
\end{align}
Equation (\ref{eq:Hasphi}) can be solved in terms of the Lambert
W-function, $W\left( z\right)$,  which is defined by the equation
\begin{equation}\label{eq:lambert}
W\left( z\right) \mbox{e}^{W\left( z \right)} = z .
\end{equation}
Eq.\ (\ref{eq:lambert})  matches (\ref{eq:Hasphi}) with
\begin{equation}
W\equiv 2p H^2 \,\,\, ,\,\,\, z \equiv -\frac{4!}{2 V^2 p}
\left( \partial \phi\right)^2 .
\end{equation}
The Lagrange multiplier part of the action is expressed in terms of
the Lambert function as
\begin{equation}
\partial_R \phi \,\epsilon^{RMNPQ}H_{MNPQ} = \sqrt{-g} \,V W
\mbox{e}^{\frac{1}{2}W} .
\end{equation}
So, finally, the dual action is given as
\begin{equation}\label{eq:dualac}
S_H = \int d^5 x \sqrt{-g} \left\{ -V\left( 1 - W\right)
  \mbox{e}^{\frac{W}{2}}\right\} ,
\end{equation}
where $W$ is the Lambert function and its argument
\begin{equation}
z=  - \frac{4!}{2V^2p} \left( \partial \phi\right)^2
\end{equation}
is the usual kinetic term in a conventional scalar Lagrangian.

Now, we are going to expand the dual action around a classical
solution. The classical solution can be read off from our dual
solution
\begin{equation}
z_0 = -\frac{4!f^2}{2V^2 p}\left( \phi_0^\prime\right)^2 = W_0 \mbox{e}^{W_0} =
-\frac{2pQ }{a^8}\mbox{e}^{-\frac{2pQ}{a^8}} .
\end{equation}
In  terms of scalar fluctuations, $\delta \phi=\phi- \phi_0$, one gets
\begin{equation}
z- z_0 = -\frac{4! f^2}{ V^2 p}\phi_0^\prime \delta
\phi^\prime-\frac{4!}{2V^2p} \left( f^2 \left(\delta
    \phi^\prime\right)^2 + e^{-2}\eta^{\mu\nu}\partial_\mu \delta
  \phi \partial_\nu \delta \phi\right) +\ldots
\end{equation}
where dots stand for terms of higher order in fluctuations. It is
useful to note that there is a term quadratic in fluctuations in
\begin{equation}
\left( z - z_0\right)^2 = \frac{(4!)^2f^4}{V^4 p^2 }\left( \phi_0
  ^\prime\right)^2 \left( \delta \phi^\prime\right)^2 +\ldots ,
\end{equation}
and all higher powers in $z-z_0$ contain higher order terms in
fluctuations.

To determine the action quadratic in scalar fluctuations we Taylor
expand the action (\ref{eq:dualac}) around $z=z_0$ till second
order. To this end, we need the $z$-derivatives of $W$ at
$z=z_0$. These can be obtained from differentiating the defining
equation (\ref{eq:lambert}) w.r.t. $z$. In order to avoid confusion
with $r$-derivatives we denote a $z$-derivative by a dot and place a
subscript zero when the argument is $z_0$. We obtain
\begin{align}
\dot{W} = \frac{1}{1+W}\mbox{e}^{-W}  &,\,\,\, \ddot{W} =
  -\frac{2+W}{\left( 1+W\right)^3}\mbox{e}^{-2W} ,\nonumber\\
\frac{d}{dz} \left[ \left( 1-W\right) \mbox{e}^{\frac{W}{2}}\right] =
-\frac{1}{2} \mbox{e}^{-\frac{W}{2}} &,\,\,\,  \frac{d^2}{dz^2} \left[ \left(
    1-W\right) \mbox{e}^{\frac{W}{2}}\right] = \frac{1}{4\left( 1+W\right)
}\mbox{e}^{-\frac{3}{2} W}  .
\end{align}
Collecting everything, we obtain for the action quadratic in scalar
fluctuations
\begin{equation}
S^{(2)} = -\frac{3!}{Vp}\int d^5x \sqrt{-g}\left\{
  \mbox{e}^{-\frac{W_0}{2}} a^{-2}\eta^{\mu\nu}\partial_\mu \delta
  \phi \partial_\nu \delta \phi + \frac{1}{1+W_0}
  \mbox{e}^{-\frac{W_0}{2}} f^2 \left(\delta
    \phi^\prime\right)^2\right\} .
\end{equation}
The last term gives rise to Kaluza Klein masses in an effective 4d
theory. Non tachyonic KK masses arise if
\begin{equation}\label{eq:KKpos}
W_0 > -1.
\end{equation}
This is satisfied for the bulk region given by
\begin{equation}\label{eq:stabcon}
  r_*< r < r_0\quad {\rm and}\quad r_0<r < r^2_0/r_*
\end{equation}
with $r_*\equiv (2pQ)^{1/8} $.
Then, there are two options to avoid the instability of the tachyonic KK mode.
First, we can cutoff the 5d space at $r=r_* >0$  and $r=r^2_0/r_* <
\infty$. This would
demand the introduction of additional branes at $r=r_1>r_*$ and $r=r^2_0/r_1 < r^2_0/r_*$.
In this case, we may hope to adjust the additional brane tensions with the extra volume determined by $r_*/r_0$.
Second, without introducing additional branes, the integral of the KK mass squared over the bulk  may turn out to be positive because of the cancellation between $0<r <r_*$ and $r_*<r<r_0$.
However, the very existence of the tachyonic KK modes at the short distances close to the AdS horizon would jeopardise perturbativity due to the bulk oscillation modes with frequencies smaller than $1/r_*$.

\subsection{The curved solutions}

As noted in the previous subsection, there are only curved solutions for the brane tension satisfying $|\Lambda_1|>\sqrt{6M^3_5 |\Lambda_b|}$.
For the nearby curved solutions,
we can also consider a quadratic action for the perturbation of the dual scalar. It takes the same form as (\ref{quadaction}) except that the 4D flat metric $\eta^{\mu\nu}$ is replaced by the curved one $g^{\mu\nu}(x)$ and the warped solution $f$ is given by eq.~(\ref{fsol}).
Thus, non-tachyonic KK masses arise under the same condition as eq.~(\ref{eq:KKpos}).
From eq.~(\ref{effcc}) with $H^2=-Q/r^8$, we get the brane
junction condition as
\be
\lambda M_5^3 =
\frac{r^2_0}{6}\left[\frac{\Lambda^2_1}{6M_5^3}-|\Lambda_b|+ V
  e^{-pQ/r^8_0} \right], \label{effcos1}
\ee
where we have incorporated the dependence on the five dimensional
Planck mass $M_5$. This is obtained by multiplying the 5d Ricci scalar
by $M_5^3$ in (\ref{eq:action}).
So, as $r_0$ and $Q$ parameters are arbitrary, the 4D effective
cosmological constant $\lambda$ is undetermined.

From the stability conditions on the nearby curved solutions, we can cut off the bulk  by putting an additional brane with tension $\Lambda_2$ at  $r=r_*$ with $r_*=(2pQ)^{1/8}$ and restrict ourselves to the region, $r_*<r<r_0$ or $r_0<r< r^2_0/r_*$.
Therefore, similarly to eq.~(\ref{effcos1}),  the junction condition on the second brane is
\be
\lambda M_5^3 =\frac{r^2_*}{6}\left[\frac{\Lambda^2_2}{6M_5^3}-|\Lambda_b|+ V
  e^{-pQ/r^8_*}
\right]=\frac{(2pQ)^{1/4}}{6}\left[\frac{\Lambda^2_2}{6 M_5^3}-|\Lambda_b|+
  V e^{-\frac{1}{2}}\right].
\label{eq:matchstar}
\ee
Our strategy to investigate the amount of fine tuning needed to
satisfy (\ref{effcos1}) and (\ref{eq:matchstar}) will be as
follows. We solve (\ref{effcos1}) by selftuning, i.e.\ by adjusting
the integration constant $Q$. The right hand side of
(\ref{eq:matchstar}) yields then the effective cosmological constant
$\lambda$ for which we impose an upper bound given by observations. To
be able to solve (\ref{effcos1}) without too much initial fine-tuning
$r_0$ should be such that
\dis{ \left| 1 - e^{-pQ/r_0^8}\right| \gsim \frac{1}{100} .\label{eq:bulktune1}
}
This yields a relation between observer brane's and cutoff brane's
postions
\begin{equation}\label{eq:zerostar}
r_0^2 \sim 2 r_\star^2 .
\end{equation}
The observational bound on $\lambda$ is
\begin{equation}
\lambda \lsim 10^{-120} M_4 ^2 \lsim  10^{-120} M_5^{\frac{9}{2}}\left( r_0^2 -
  r_\star^2\right) \sqrt{\frac{6}{- \Lambda_b - V e^{-pQ/r_\star^8}}},
\end{equation}
where $M_4$ denotes the 4d effective Planck mass and the second
inequality is obtained as in (\ref{eq:effplanck}). Plugging that into
(\ref{eq:matchstar}) we obtain ($(r_0^2 -r_\star^2)/r_\star^2 \sim 1$ because of
(\ref{eq:zerostar}))
\begin{equation}\label{eq:bulktune}
10^{-120} \gsim M_5^{-\frac{15}{2}}\left(
  \frac{\Lambda_2^2}{6M_5^3} -|\Lambda_b| + V
  e^{-\frac{1}{2}}\right)\sqrt{-\Lambda_b -V e^{-\frac{1}{2}}} .
\end{equation}
Without fine-tuning there are no major cancellations among
contributions on the right hand side of
(\ref{eq:bulktune}). Therefore, we get conditions for each of the parameters
\begin{equation}\label{eq:bulkfine}
\frac{\Lambda_2^2}{M_5^3} \,\,\, ,\,\,\, \left| \Lambda_b\right|\,\,\,
,\,\,\, \left| V\right| \lsim 10^{-80} M_5^5 .
\end{equation}
Note, that since we imposed an initial tuning of $1/100$ in
(\ref{effcos1}) the same bound appears also for $\Lambda_2$ replaced
by $\Lambda_1$. Condition (\ref{eq:bulkfine}) means either that
quantities on the left hand side are composed of finely tuned
cancelling contributions or that there is some symmetry above a scale
\begin{equation}\label{eq:fivep}
M_s\sim 10^{-16} M_5
\end{equation}
protecting these quantities against quantum corrections. Such a
symmetry might, for instance, be supersymmetry. In any case, experiment
provides a lower bound on its breaking scale
\begin{equation}
M_s > 1 \mbox{TeV}.
\end{equation}
In what follows we take $M_s$ to be a TeV (anything above would
increase the amount of fine-tuning to be found shortly). With our
findings so far we obtain for the 4d effective Planck mass
\begin{equation}
M_4 ^2 \sim M_5^2 r_\star^2 \, 10^{40}\sim M_s ^2 r_\star^2 10^{72} .
\end{equation}
With the symmetry breaking scale at about a TeV we find for the
brane's position
\begin{equation}\label{eq:braneposi}
r_0^2 \sim r_\star^2 \sim 10^{-40} .
\end{equation}
It remains to check the first of the conditions in
(\ref{eq:bulkfine})
\begin{equation}\label{eq:branefine}
\Lambda_i^2 < 10^{-80} M_5 ^8 ,
\end{equation}
where $i=1,2$. Let's focus on the observer's brane
at $r_0$. Assuming broken supersymmetry at $M_s$ we can parameterise
fine-tuning by a number $\alpha$ as follows
\begin{equation}
r_0^4 \Lambda_1 = \alpha M_s^4 ,
\end{equation}
where on the left hand side the vaccuum energy on the brane measured
in Minkowski frame (i.e.\ canonical kinetic terms) appears. Hence,
no fine-tuning corresponds to $\alpha \sim 1$. However, condition
(\ref{eq:branefine}) together with (\ref{eq:fivep}) and
(\ref{eq:braneposi}) yields
\begin{equation}
\alpha \sim 10^{-56} .\label{eq:finDegree}
\end{equation}
So, the amount of fine-tuning needed is about the same as in
conventional 4d supersymmetric theories.

Once, we realize that  we need severe fine-tuning we might as well increase the severeness of the bulk tuning condition (\ref{eq:bulktune1}). That step is motivated by the hope of getting less severe conditions on the brane parameters at the price of increasing the number of fine-tuning conditions. We consider three cases parametrized as follows
\dis{
\left| 1-e^{-pQ/r_0^8}\right|\ge \alpha^b,~~ r_0^4\Lambda_1= \alpha^a M_s^4,
}
with
\dis{
(a,b)= \left(1,\frac{1}{28} \right),~ (1,1),~(0,1).
}
The first case is the one we considered above. It is characterized by severe fine-tuning of the brane parameters and only mild tuning of bulk parameters. (No tuning of the bulk parameters would correspond to $b=0$.) In the second case, fine-tuning is distributed equally between brane and bulk parameters, and in the last case fine-tuning is imposed only on bulk parameters. Performing an analysis similar to the one we carried out for the first case we obtain
\dis{
\alpha\lesssim 10^{-\frac{112}{2a+b/2}}
}
reproducing (\ref{eq:finDegree}) for the first case. For the second case of equally distributed fine-tuning we find
\dis{
\alpha\lesssim 10^{-45}.
}
So, indeed the amount of fine-tuning can be reduced by distributing it
over bulk and brane parameters. The third case, with fine-tuning of
bulk parameters only, leads to the most severe condition
\dis{
\alpha\lesssim 10^{-224}.
}

\section{General discussion and no-go theorem}

In this section, we discuss stability conditions for the general form of the action with a three-form field in 5d and show that those conditions are incompatible with the self-tuning condition.

Following a similar procedure as in the previous section, for a general form of the bulk action with $K(H^2)$, we get the dual scalar action as
\be
S_{\rm dual}=-\int d^5x \sqrt{-g}\, G(W)   \label{gendualaction}
\ee
where we find that the following $G(W)$ is useful in deriving the no go theorem,
\be
G(W)\equiv 2W \frac{\partial K}{\partial W}-K(W),\quad  W\equiv H^2<0,
\ee
and the dual scalar is given by
\be
z\equiv - c  (\partial\phi)^2= \frac{c}{3!}\Big(\frac{\partial K}{\partial W}\Big)^2 W \label{dualscalar}
\ee
where a constant parameter $c$ is chosen to be positive without loss of generality. Then, expanding the dual scalar around the background solution as $\phi=\phi_0+\delta\phi$, we obtain the quadratic action for the scalar perturbation as
\be
S_{\rm dual}\simeq \int d^5x\sqrt{-g} \bigg[c \,a^{-2} \eta^{\mu\nu} (\partial_\mu \delta\phi) (\partial_\nu\delta \phi)\,\frac{\partial G}{\partial z}\Big|_{z=z_0}+c f^2 \Big(2z \frac{\partial^2 G}{\partial z^2}+\frac{\partial G}{\partial z}\Big)\Big|_{z=z_0}(\delta\phi')^2   \bigg].  \label{quadaction}
\ee
Thus, the conditions for neither ghost or tachyonic KK modes are
\be
\frac{\partial G}{\partial z}\Big|_{z=z_0}<0,\quad \quad \Big(2z \frac{\partial^2 G}{\partial z^2}+\frac{\partial G}{\partial z}\Big)\Big|_{z=z_0}<0. \label{goodcond}
\ee
Using the derivatives as follows,
\bea
\frac{\partial G}{\partial z}&=& -J(W)\frac{\partial G}{\partial W}, \\
\frac{\partial^2 G}{\partial z^2}&=& J(W)\Big(\frac{\partial J}{\partial W}\frac{\partial G}{\partial W}+J\frac{\partial^2 G}{\partial W^2}\Big), \\
\frac{\partial G}{\partial W}&=&\frac{\partial K}{\partial W}+2 W\frac{\partial^2 K}{\partial W^2} \label{Gder}
 \eea
where use is made of eq.~(\ref{dualscalar}) and
\be
J(W)\equiv-\frac{\partial W}{\partial z}= -\frac{3!}{c}\Big(\frac{\partial K}{\partial W}\Big)^{-1}\Big[\frac{\partial K}{\partial W}+2 W\frac{\partial^2 K}{\partial W^2} \Big]^{-1},
\ee
we rewrite the conditions given in eq.~(\ref{goodcond}) as
\bea
&&A\equiv -\frac{3!}{c}\Big(\frac{\partial K}{\partial W}\Big)^{-1} >0, \\
&&B\equiv \frac{3!}{c}\Big(-1+2z_0 \frac{\partial J}{\partial W}\Big)\Big(\frac{\partial K}{\partial W}\Big)^{-1}-2z_0J^2 \frac{\partial^2 G }{\partial W^2} >0. \label{2nd}
\eea
The second condition (\ref{2nd}) is further simplified to
\be
B=A\Big[1-2W\frac{\partial^2 K}{\partial W^2}\Big(\frac{\partial G}{\partial W}\Big)^{-1}\Big]=-\frac{3!}{c}\Big(\frac{\partial G}{\partial W}\Big)^{-1}>0. \label{2nda}
\ee
where use is made of $2W\frac{\partial^2 K }{\partial W^2}=-\frac{\partial K}{\partial W}+\frac{\partial G}{\partial W}$ in the second equality.
Eventually, independent of the sign of $\frac{\partial K}{\partial W}$, we get the second condition (\ref{2nd}) as
\be
\frac{\partial G}{\partial W}<0.
\ee
We note that from $A>0$, the function $K(W)$ must be a monotonically decreasing function of $W$ for $W<0$. On the other hand, whether or not $B>0$ is satisfied depends on the detailed form of the function $K$.

Together with the criterion for finite Planck mass (\ref{stcond}),
we now collect the necessary conditions for stable self-tuning solutions,
\bea
&&1)\quad |W|^{-1/4} K(W)\rightarrow 0\quad \quad{\rm for}\,\,\, W\rightarrow -\infty, \label{selftuning}\\
&& 2)\quad \frac{\partial K}{\partial W} <0,  \label{2ndaa}\\
&& 3)\quad \frac{\partial G}{\partial W}=\frac{\partial K}{\partial W}+2 W\frac{\partial^2 K}{\partial W^2}<0 \label{3rdaa}
\eea
where use is made of $W=H^2=-Q/a^8$ in the first line. The first self-tuning condition (\ref{selftuning}) fixes the asymptotic behavior of the function $K(W)$ at $W=-\infty$ to have a power less than $\frac{1}{4}$.

Let's take some examples.
For instance, for $K(H^2)=1/H^2=1/W$ \cite{kkl}, we get $\frac{\partial K}{\partial W}=-1/W^2$, $G(W)=-3/W$ and $\frac{\partial G}{\partial W}=3/W^2>0$. Thus, this example leads to tachyonic KK modes. In general, for the form, $K=A(-W)^{\alpha/4}$ with $A>0$ and $\alpha$ being a constant parameter, the self-tuning condition (\ref{selftuning}) leads to $\alpha <1$.
On the other hand, from the positivity of the kinetic term, (\ref{2ndaa}), we get $\alpha>0$, resulting in $0<\alpha < 1$ with the self-tuning condition.
But, the non-tachyonic condition (\ref{3rdaa}) requires $\alpha<0$ or $\alpha>2$, so it is not consistent with the self-tuning solution with no ghost.

As the second example, we take $K(H^2)=-e^{H^2}=-e^W$ \cite{generalST} so we get $\frac{\partial K}{\partial W}=-e^W$, $G(W)=(1-2W)e^W$ and $\frac{\partial G}{\partial W}=-(1+2W) e^W$. Then, this does not give tachyonic modes only for $W>-\frac{1}{2}$.
For another example with $K(H^2)=\log(-H^2)$ \cite{generalST}, we get $\frac{\partial K}{\partial W}=-\frac{1}{W}>0$ and $\frac{\partial G}{\partial W}=\frac{1}{W}<0$, so there are ghost modes in the case.

Now we prove a no-go theorem. Condition (\ref{3rdaa}) can be written
as (prime denotes derivative w.r.t.\ $W$) 
\begin{equation}\label{eq:mono}
\left( \sqrt{-W}K^\prime\right)^\prime > 0
\end{equation}
This implies that $\sqrt{-W}K^\prime$  is monotonically increasing or,
since $K^\prime < 0$, that $\sqrt{-W}\left|K^\prime\right| \geq 0$  is
monotonically decreasing. This implies that
\begin{equation}\label{eq:contasy}
\sqrt{-W}\left|K^\prime\right|> 0, \,\,\,\mbox{for}\,\,\, W\to -\infty ,
\end{equation}
On the other hand, taking $W\frac{\partial}{\partial W}$ of (\ref{selftuning})
yields
\begin{equation}\label{eq:kprime}
(-W)^{\frac{3}{4}} K^\prime \to 0 \,\,\,\mbox{for}\,\,\, W\to -\infty  .
\end{equation}
So, from (\ref{eq:contasy}) and (\ref{eq:kprime}), self-tuning and stability seems not compatible.

%
%

Let us consider two cases. The first case is that the second term of
(\ref{quadaction}) has the wrong sign for some region in the bulk
whereas the first term has always the correct sign. This case has been
encountered at the end of section 3.1 and the problem has been discussed in section 3.2.

The other potentially interesting case is that the second term in
(\ref{quadaction}) has always the correct sign. Then $r$ dependence in
fluctuations is suppressed. We can integrate over $r$ and hope that
the effective kinetic term for the fluctuations turns out with the
correct sign.
As discussed before, the tachyon-free condition  is given by eq.~(\ref{3rdaa}) or (\ref{eq:mono}) through the bulk, independent of whether the kinetic term is of correct sign or not.
From our previous discussion, it is clear that $K$ must have the wrong sign for $W\to
-\infty$ ($r\to 0$) (to avoid conclusion (\ref{eq:contasy})). We also need
a region in which $K^\prime$ is negative to have contributions
potentially cancelling the ones with the wrong sign.
First, we assume that $K^\prime$ is continuous. Then there must be
some position $W_0$ where $K^\prime \left( W_0\right) = 0$. Further,
$W_0 < 0$ since otherwise the brane at $r= r_0$ would be localised at
infinity.
Evaluating (\ref{eq:mono}) at $W_0$ yields
$$ K^{\prime\prime}\left( W_0\right) > 0 . $$
That means $K$ should have a local minimum there. This contradicts our
assumption that $K$ changes from a monotonically increasing to a
monotonically decreasing function at $W_0$. So, $K^\prime$ cannot be
continuous. Next, we assume that $K^\prime$ has a removable singularity at
$W_0$, i.e.\ it is finite and jumps by a finite amount from positive
to negative. Then we would have a $\delta$-function contribution with
a negative sign to (\ref{eq:mono}). Fluctuations localised at $W_0$
would be destabilised. So, $K^\prime$ must diverge at some point
$W=W_0$ which implies that $K$ is not continuous or diverges at
$W=W_0$. That would imply that our metric degenerates at $W=W_0$ and
we would need to cut off the extra dimension above that point. An
example is
\begin{equation}
K= \log\left| \frac{1+ \sqrt{-W}}{1-\sqrt{-W}}\right| .
\end{equation}

\section{Conclusions}

We have studied stability conditions for the self-tuning solutions with a brane in 5d gravity with the addition of a three form gauge potential.  In this model, unorthodox bulk Lagrangians give rise to a general class of the self-tuning solutions where the 4d effective cosmological constant is adjusted by the change of integration constants, satisfying the absence of a naked singularity and the finiteness of the 4d Planck mass.  We have performed the perturbation analysis for the 4d effective theory of such self-tuning solutions and have proved the no go theorem that there always exist tachyonic KK masses or ghosts for the self-tuning solutions with any continuous form of the bulk Lagrangian in this model. 
Choosing an exponential type of the bulk Lagrangian as an example, we
showed that there is a KK tachyonic instability in some region of the
bulk space and an extra brane introduced to cure the instability
problem leads to a fine-tuning being as severe as in 4d. 
Therefore, a totally new idea would be required to circumvent the no go theorem and it would bring us one step further in understanding the cosmological constant problem in terms of self-tuning solutions.

\section*{Acknowledgments}
The work of S.F. was partially supported by the SFB-Tansregio TR33
``The Dark Universe'' (Deutsche Forschungsgemeinschaft) and the European Union 7th network program ``Unication in the LHC era'' (PITN-GA-2009-237920).
The work of J.E.K. is supported in part by the National Research Foundation (NRF) grant funded by the Korean Government (MEST) (No. 2005-0093841).

\end{document}